# Proximity Effect of Epitaxial Iron Phthalocyanine Molecules on High-Quality Graphene Devices

Haiyang Pan[1*], Xiaobo Wang[2*], Qiaoming Wang[1], Xiaohua Wu[1], Chang Liu[1], Nian Lin[2†], and Yue Zhao[1,3‡]

[1]Department of Physics, Southern University of Science and Technology, Shenzhen 518055, China

[2]Department of Physics, The Hong Kong University of Science and Technology, Hong Kong, China

[3]Shenzhen Institute for Quantum Science and Engineering, Southern University of Science and Technology, Shenzhen 518055, China

[*]H. Pan and X. Wang contributed equally to this work.
[†]Corresponding author: phnlin@ust.hk
[‡]Corresponding author: zhaoy@sustech.edu.cn

**Abstract**

Depositing magnetic insulators on graphene has been a promising route to introduce magnetism via exchange proximity interaction in graphene for future spintronics applications. Molecule-based magnets may offer unique opportunities because of their synthesis versatility. Here, we investigated the magnetic proximity effect of epitaxial iron phthalocyanine (FePc) molecules on high-quality monolayer and bilayer graphene devices on hexagonal boron nitride substrate by probing the local and non-local transport. Although the FePc molecules introduce large hole doping effects combined with mobility degradation, the magnetic proximity gives rise to a canted antiferromagnetic state under a magnetic field in the monolayer graphene. On bilayer graphene and FePc heterostructure devices, the non-local transport reveals a pronounced Zeeman spin-Hall effect. Further analysis of the scattering mechanism in the bilayer shows a dominated long-range scattering. Our findings in graphene/organic magnetic insulator heterostructure provide a new insight for the use of molecule-based magnets in two-dimensional spintronic devices.

Graphene possesses unusual electronic properties of high carrier mobility, long spin relaxation distance, and lifetime, providing an attractive prototype gate tunable platform for the manipulation of spin towards the next-generation spin-based logic devices.[1-3] The generation of spin-polarized current is usually initialized by spin injection from ferromagnetic electrodes, where the contact interfaces could limit the spin injection efficiency.[3] The unique band structure of graphene allows alternative approaches to creating spin transport without magnetic contacts by introducing magnetism in graphene.[4] The strong hybridization of graphene and magnetic materials can generate proximity-induced spin polarization and exchange splitting to the electronic structure of graphene. Direct adsorption of magnetic atoms or molecular magnets is expected to introduce proximity-induced magnetism effects in graphene.[5-8] However, the magnetic dopants can also act as impurities scattering centers, resulting in mobility degradation and doping effects, thus impacting electrical transport performance.[9-12] Bringing graphene in contact with ferromagnetic insulators has been considered a promising step to introduce a magnetic exchange field for spin-polarized edge conduction. Magnetic proximity effects induced strong Zeeman splitting have been observed in several graphene/magnetic insulator heterostructure systems, including EuS films, $CrBr_3$ thin layers, and $BiFeO_3$ nanoplates.[4, 13-15]

Despite the recent progress, there has been little investigation of graphene heterostructure devices with molecular-based magnets. Transition metal phthalocyanines have been known as organic magnetic semiconductors where their magnetic properties can be sensitive to the crystal phase of the $\pi-$conjugated polymers, depending on the metal ion and orbital overlap.[16] They are cost-effective and easily procurable organic materials. The magnetic molecules exhibit localized electronic spin states with long coherence time and are thus suitable for hybridization with nonmagnetic films to induce spin transport in versatile two-dimensional organic spin interface architectures.[17, 18-24] Driven by the $\pi-\pi$ interaction, the polymers can form stable crystalline phase on graphene, offering a integrated graphene/organic ferromagnet interface for the investigation of magnetic proximity effect.

Previous work demonstrated that large-scale ordered flat islands of iron (II) phthalocyanine (FePc) molecules could be formed on graphene when sub-monolayer amounts of molecules were deposited.[19] The epitaxial film stabilized by intermolecular non-covalent interactions with the graphene surface can form periodic square patterns of FePc molecules with a lattice constant of 1.4 nm. The absence of inhomogeneous cluster formation also mitigates additional disorder scattering in the transport properties of graphene. The bulk of FePc molecules exhibits ferromagnetic order with Tc around 10 K.[22] For FePc thin films, its properties are related to the deposition substrate. The magnetization measurements and X-ray magnetic circular dichroism results reveal that the FePc thin films on gold surface exhibit ferromagnetic order in the molecule plane below 4.5 K.[16]

In the present study, we investigate the magnetic proximity effect of epitaxially grown FePc film on monolayer and bilayer graphene by comparing the local and non-local transport measurements. To minimize charge impurity scattering and improve carrier mobility, we transfer graphene on hexagonal boron nitride (h-BN) for high-quality electrical properties.[20, 21] Our measurements show that the magnetic proximity effect from FePc deposition creates different ground state on monolayer and bilayer graphene devices. On monolayer graphene, the non-local measurements showed a suppressed peak near the charge neutrality point under a magnetic field, while the local transverse resistance remains high, indicating the formation of a canted antiferromagnetic (CAFM) state from the magnetic proximity effect. On bilayer graphene, we observe a reducing local resistance and an increasing non-local resistance at the Dirac point as the magnetic field increases, consistent with the formation of a Zeeman spin Hall effect. By analyzing the ratio of transport scattering time ($\tau_t$) to quantum scattering time ($\tau_q$) in bilayer graphene/FePc heterostructure, we believe that the transport properties are dominated by long-range charge impurity scattering.

**RESULTS AND DISCUSSION**

Figure 1a shows the molecular structure of FePc, where one Fe atom is connected with four pyrrolic N atoms in square-planar coordination. These planar molecules formed linear chains by columnar stacking in the thin films. The monolayer of FePc molecules deposited on the graphene surface formed a well-ordered flat array. Figure 1c shows the scanning tunneling microscopy (STM) topography image of a FePc monolayer on HOPG. The planes of FePc molecules are parallel to the substrate, similar to the structure of FePc on a gold substrate.[16] After verifying the array-order of our FePc film, more than one layer of FePc molecules are deposited onto the surface of the G/BN devices to ensure the coverage and the magnetic interface. The deposition of FePc was performed at an ultra-high vacuum environment with in-situ vacuum annealing of the G/BN devices so that the atomic contact at the interface can be guaranteed.

Figure 1d presents the gate voltage ($V_g$) dependent resistance for a typical monolayer graphene device before and after FePc deposition. After the FePc deposition, the monolayer graphene device experiences a rightwards shift of charge neutrality point and a broadened Dirac peak, implying a hole doping effect and impurity scattering by the organic molecule. The FePc film introduces an initial hole doping on the level of $8.7 \times 10^{11}$ cm$^{-2}$. The mobility, obtained by fitting the relation between conductivity ($\sigma$) and the carrier density ($n$) (Figure S3), decreases by one order from 139, 000 cm$^2$V$^{-1}$s$^{-1}$ to 14, 000 cm$^2$V$^{-1}$s$^{-1}$. Charge inhomogeneity $\delta n$ calculated from the width of the Dirac peak increases from $7.0 \times 10^{10}$ cm$^{-2}$ to $4.8 \times 10^{11}$ cm$^{-2}$. Similar hole doping and mobility degradation were observed in bilayer devices, as shown in Figure S2. In addition, the conductivity curves $\sigma(n)$ for both the monolayer and bilayer devices (Figure S3) exhibit behavior change from sublinear to linear after FePc deposition, consistent with a dominant charge impurity scattering caused by the inhomogeneous charge doping from the adjacent FePc molecules.[10, 22-25] Density functional theory (DFT) calculations on the band structure of the FePc/monolayer graphene (FePc/MG) heterostructure (Figure S4) shows an additional state contributed from the Fe 3$d$ orbital near the Dirac cone at the $\Gamma$ point, which may contribute to the electron-hole asymmetry behavior of the monolayer graphene device after the FePc deposition.

When the magnetic order is formed in the FePc film at T < T$_c$, the strong exchange field is expected to result in a Zeeman splitting effect in graphene, as shown in Figure 2a. The spin degeneracy at the charge neutrality point can be lifted, leading to sub-bands with opposite spin due to the Zeeman splitting. When the Fermi level lies between the sub-band Dirac points, both spin-up holes and spin-down electrons exist. Under external magnetic field, as illustrated in Figure 2b, an enhanced non-local signal can be observed near the Dirac point due to the Zeeman splitting.[13, 14, 26-28]

Figure 2c presents the gate dependent non-local resistance of the FePc/MG device under magnetic field at T = 1.6 K. A dip in the non-local resistance appears near the Dirac point when the magnetic field approaches 8 T. As increasing magnetic field, this dip becomes more prominent with an almost vanishing non-local resistance. On the other hand, the local transport measurement shows that the local resistance at the Dirac peak increases with an external magnetic field and tends to saturate at B > 7 T (Figure 2e). The non-local resistance dip and saturated local resistance at the Dirac peak are in sharp contrast to the formation of gapless spin-polarized edge channels due to the Zeeman spin Hall effect, where an increase of non-local resistance and a decrease of local resistance with magnetic field are expected. Such local and non-local behaviors near Dirac point can be explained by a canted antiferromagnetic (CAFM) ground state when there is a competition between the valley anisotropy and Zeeman splitting energy. As shown in the schematic diagram of the CAFM state (Figure 2d), the opening of an edge state gap can effectively decrease the non-local signal and enhance longitudinal resistance at the Dirac point ($R_{xx,D}$). There is a saturation behavior of $R_{xx,D}$ at B > 7 T, possibly due to the disorder scattering from the neighboring FePc molecules.

We then investigate the proximity effect of FePc film on bilayer graphene. Figure 3a and 3b show the non-local and local curves of a bilayer graphene/FePc (FePc/BG) heterostructure device. As summarized in Figure 3c, when the magnetic field increases, the non-local peak value $R_{NL,D}$ keeps increasing from 0 to 10 T, while the local resistance peak increases first (0 < B < 2 T ), then quickly decreases (2 T < B < 10 T). The associated local and non-local transport behaviors are consistent with a ground state ($v$ = 0) phase of an FM state because of the magnetic proximity effect from FePc film. Figure 3d shows the schematics of the FM phases for bilayer graphene based on the quantum Hall ferromagnetism theory.[29-31] Due to the extra orbital degeneracy of the zero-energy Landau level in bilayer graphene, the number of branches is doubled compared to monolayer graphene. We focus mainly on the evolution of spin-polarized edge channels. The edge gap closes at the FM phase, and there exist gapless counter-propagating edge channels with opposite spins, leading to a decrease of $R_{xx,D}$ and an increase of $R_{NL,D}$. We note that local peak value $R_{xx,D}$ increases first from 0 T to 2 T before the FM phase emerges. The onset of FM phase at 2 T may be related to the hole doping effect of FePc film, which could induce orbital splitting because of the displacement field, leading to a competition between the Zeeman splitting and orbital splitting at lower magnetic field.[32] The analysis of ohmic contribution (Figure S6) shows negligible ohmic leakage compare with the non-local signal. The temperature-dependent $R_{NL,D}$ at 4 T (Figure S7) shows a transition around 15 K, which is very close to the $T_c \sim 10$ K of bulk FePc (our final deposition thickness is about half a micron), confirming the magnetic proximity effect from the FePc film.

To further elucidate the influence of the FePc film on the electron transport properties of bilayer graphene, we conduct Shubnikov-de Haas (SdH) oscillation analysis to reveal the scattering mechanism. The magnetoresistance transport results of the original bilayer graphene are displayed in Figure 4a. A prominent SdH oscillation could be observed at various gate voltages, where the oscillation frequency can be obtained by fast Fourier transform (FFT). Figure 4b displays the magnetoresistance results of the FePc/BG device. The carrier-density-dependent frequency curves (Figure S10) were almost identical before and after FePc deposition, indicating the negligible influence of FePc on the Fermi surface size.

The effective mass can be extracted (Figure S9) from the SdH oscillations by fitting the formula:[1]

$$\frac{A(T)}{A(T_0)} = \frac{T\sinh(2\pi^2 k_B T_0 m^*/\hbar e B_n)}{T_0\sinh(2\pi^2 k_B T m^*/\hbar e B_n)}$$

where $A(T)$ and $A(T_0)$ are the FFT amplitudes of the SdH oscillations at temperatures $T$ and $T_0$, respectively, and $B_n$ is the magnetic field corresponding to the peaks of the oscillations. The quantum scattering time ($\tau_q$) and transport scattering time ($\tau_t = m^*\mu/e$), before and after the FePc deposition, can be determined, as shown in Figure 4c-d. Calculation details of $\tau_q$ can be found in the supplementary material. The $\tau_q$ of the original graphene ranged from 80 to 130 fs, larger than the reported value in the bilayer graphene on a $SiO_2$ substrate.[33] The $\tau_q$ slightly decreased to 40 -100 fs after FePc deposition. The transport scattering time $\tau_t$ also decreased from ~500 fs to less than 200 fs after the deposition.

The ratio of $\tau_t/\tau_q$ can be a powerful indicator of the complex scattering sources in 2D electron systems.[33] While $\tau_t/\tau_q$ ~1 indicates wide-angle scattering (related to the short-ranged scattering source), a larger $\tau_t/\tau_q$ could result from small-angle scattering (related to the long-range charged impurity scattering). Figure 4e displays $\tau_t/\tau_q$ ratio of the bilayer graphene device before and after FePc deposition. Before FePc deposition, the $\tau_t/\tau_q$ values ($\tau_t/\tau_q > 4$ on the hole side and $1 < \tau_t/\tau_q < 3$ on the electron side) were in agreement with the reported values for graphene on the $SiO_2$ substrate, suggesting long-range Coulomb disorder scattering originated from the graphene/substrates interface. After FePc deposition, however, $\tau_t/\tau_q$ values showed a significant decrease, ranging from 0.9 and 1.7. Such a decrease is related to the increased short-range scattering from the magnetic impurities induced by FePc molecules.

**CONCLUSIONS**

In conclusion, we have systematically investigated local and non-local transport properties of monolayer and bilayer graphene devices with epitaxial FePc films. The magnetic proximity effect from FePc film results in a Zeeman splitting field. A canted antiferromagnetic ground state was observed on monolayer graphene, and a pronounced Zeeman Hall effect was realized on bilayer graphene. Our results reveal the magnetic proximity effect at the interface of graphene/organic magnetic insulator heterostructures and provide a new insight for using molecule-based magnets in two-dimensional spintronic devices.

**EXPERIMENTAL METHODS**

The graphene/h-BN (G/BN) samples were prepared using the dry transfer method.[34-36] Graphene pieces were obtained by mechanical exfoliation on polymer methyl methacrylate (PMMA) substrate. h-BN flakes with thicknesses of 20-40 nm were exfoliated on oxidized Si wafers. Mono and bilayer graphene pieces were identified by initial optical microscopy and subsequent Raman spectroscopy. The selected graphene pieces were dry-transferred onto the target h-BN flakes using a transfer station. The G/BN heterostructures were then annealed in an argon and hydrogen atmosphere at 350 °C for three hours to remove the polymer residue.[36] We then selected the graphene areas free from bubbles and wrinkles by atomic force microscopy, where standard electron beam lithography (EBL) and metal evaporation were performed for standard Hall-bar devices. Secondary EBL and reactive ion etching were employed to etch the rest part. The final devices were reannealed in a mixed argon and hydrogen atmosphere to remove the resist residual.

The FePc molecules purchased from Sigma-Aldrich with the dye content 90% were purified in the ultra-high vacuum (UHV) chamber for several cycles through heating. Those molecules were then loaded into an organic molecular beam epitaxy (OMBE) system and out-gassed for several hours before deposition. The G/BN devices were transferred into the OMBE chamber and annealed at 220 °C for 10 hours to remove any organic and moisture residue. The deposition of FePc film occurred at an evaporation temperature of 380 °C, while the devices were maintained at 135 °C to optimize the crystallite size. STM characterization was carried out in a CreaTec UHV STM system working at 78 K. The growth rate was 1.1 Å/s, and the estimated thickness of the final film was approximately 462 nm.

The transport measurements were performed in a He$^4$ cryostat system (Oxford Equipment). The four-probe local and non-local measurements were performed using a standard lock-in technique with an AC excitation current of 100 nA. A Keithley source meter (K2400) was applied to the bottom of the SiO$_2$ (285 nm)/ *p*-Si substrate as the gate voltage.

## Acknowledgments

The research was supported by the National Natural Science Foundation of China under project No.11674150, the Key-Area Research and Development Program of Guangdong Province 2019B010931001, Guangdong Innovative and Entrepreneurial Research Team Program 2016ZT06D348. The authors would like to acknowledge the technical support from the SUSTech CRF. XBW and NL thank the financial support from Hong Kong RGC 16300617 and C6012-17E.

## Notes

The authors declare no competing financial interest.

## References

1. Novoselov, K. S.; Geim, A. K.; Morozov, S. V.; Jiang, D.; Katsnelson, M. I.; Grigorieva, I. V.; Dubonos, S. V.; Firsov, A. A., Two-dimensional gas of massless Dirac fermions in graphene. *Nature* **2005,** *438* (7065), 197-200.
2. Zhang, Y.; Tan, Y.-W.; Stormer, H. L.; Kim, P., Experimental observation of the quantum Hall effect and Berry's phase in graphene. *Nature* **2005,** *438* (7065), 201-204.
3. Tombros, N.; Jozsa, C.; Popinciuc, M.; Jonkman, H. T.; van Wees, B. J., Electronic spin transport and spin precession in single graphene layers at room temperature. *Nature* **2007,** *448* (7153), 571-574.
4. Wang, Z.; Tang, C.; Sachs, R.; Barlas, Y.; Shi, J., Proximity-Induced Ferromagnetism in Graphene Revealed by the Anomalous Hall Effect. *Physical Review Letters* **2015,** *114* (1), 016603.
5. Ding, J.; Qiao, Z.; Feng, W.; Yao, Y.; Niu, Q., Engineering quantum anomalous/valley Hall states in graphene via metal-atom adsorption: An ab-initio study. *Physical Review B* **2011,** *84* (19), 195444.
6. Qiao, Z.; Yang, S. A.; Feng, W.; Tse, W.-K.; Ding, J.; Yao, Y.; Wang, J.; Niu, Q., Quantum anomalous Hall effect in graphene from Rashba and exchange effects. *Physical Review B* **2010,** *82* (16), 161414.
7. Qiao, Z.; Jiang, H.; Li, X.; Yao, Y.; Niu, Q., Microscopic theory of quantum anomalous Hall effect in graphene. *Physical Review B* **2012,** *85* (11), 115439.
8. Candini, A.; Klyatskaya, S.; Ruben, M.; Wernsdorfer, W.; Affronte, M., Graphene Spintronic

Devices with Molecular Nanomagnets. *Nano Letters* **2011,** *11* (7), 2634-2639.
9. McCreary, K. M.; Swartz, A. G.; Han, W.; Fabian, J.; Kawakami, R. K., Magnetic Moment Formation in Graphene Detected by Scattering of Pure Spin Currents. *Physical Review Letters* **2012,** *109* (18), 186604.
10. Chen, J. H.; Jang, C.; Adam, S.; Fuhrer, M. S.; Williams, E. D.; Ishigami, M., Charged-impurity scattering in graphene. *Nature Physics* **2008,** *4* (5), 377-381.
11. Chao-Yi Cai, J.-H. C., Electronic transport properties of Co cluster-decorated graphene. *Chin. Phys. B* **2018,** *27* (6), 67304-067304.
12. Qin, Y.; Wang, S.; Wang, R.; Bu, H.; Wang, X.; Wang, X.; Song, F.; Wang, B.; Wang, G., Sizeable Kane–Mele-like spin orbit coupling in graphene decorated with iridium clusters. *Appl. Phys. Lett* **2016,** *108* (20), 203106.
13. Wei, P.; Lee, S.; Lemaitre, F.; Pinel, L.; Cutaia, D.; Cha, W.; Katmis, F.; Zhu, Y.; Heiman, D.; Hone, J.; Moodera, J. S.; Chen, C.-T., Strong interfacial exchange field in the graphene/EuS heterostructure. *Nature Materials* **2016,** *15* (7), 711-716.
14. Wu, Y.-F.; Song, H.-D.; Zhang, L.; Yang, X.; Ren, Z.; Liu, D.; Wu, H.-C.; Wu, J.; Li, J.-G.; Jia, Z.; Yan, B.; Wu, X.; Duan, C.-G.; Han, G.; Liao, Z.-M.; Yu, D., Magnetic proximity effect in graphene coupled to a BiFeO3 nanoplate. *Physical Review B* **2017,** *95* (19), 195426.
15. Tang, C.; Zhang, Z.; Lai, S.; Tan, Q.; Gao, W.-b., Magnetic Proximity Effect in Graphene/CrBr3 van der Waals Heterostructures. *Adv Mater* **2020,** *32* (16), 1908498.
16. Gredig, T.; Colesniuc, C. N.; Crooker, S. A.; Schuller, I. K., Substrate-controlled ferromagnetism in iron phthalocyanine films due to one-dimensional iron chains. *Physical Review B* **2012,** *86* (1), 014409.
17. Avvisati, G.; Cardoso, C.; Varsano, D.; Ferretti, A.; Gargiani, P.; Betti, M. G., Ferromagnetic and Antiferromagnetic Coupling of Spin Molecular Interfaces with High Thermal Stability. *Nano Letters* **2018,** *18* (4), 2268-2273.
18. Gamou, H.; Shimose, K.; Enoki, R.; Minamitani, E.; Shiotari, A.; Kotani, Y.; Toyoki, K.; Nakamura, T.; Sugimoto, Y.; Kohda, M.; Nitta, J.; Miwa, S., Detection of Spin Transfer from Metal to Molecule by Magnetoresistance Measurement. *Nano Letters* **2020,** *20* (1), 75-80.
19. de la Torre, B.; Švec, M.; Hapala, P.; Redondo, J.; Krejčí, O.; Lo, R.; Manna, D.; Sarmah, A.; Nachtigallová, D.; Tuček, J.; Błoński, P.; Otyepka, M.; Zbořil, R.; Hobza, P.; Jelínek, P., Non-covalent control of spin-state in metal-organic complex by positioning on N-doped graphene. *Nature Communications* **2018,** *9* (1), 2831.
20. Dean, C. R.; Young, A. F.; Meric, I.; Lee, C.; Wang, L.; Sorgenfrei, S.; Watanabe, K.; Taniguchi, T.; Kim, P.; Shepard, K. L.; Hone, J., Boron nitride substrates for high-quality graphene electronics. *Nature Nanotechnology* **2010,** *5* (10), 722-726.
21. Zomer, P. J.; Dash, S. P.; Tombros, N.; Wees, B. J. v., A transfer technique for high mobility graphene devices on commercially available hexagonal boron nitride. *Applied Physics Letters* **2011,** *99* (23), 232104.
22. Hwang, E. H.; Adam, S.; Sarma, S. D., Carrier Transport in Two-Dimensional Graphene Layers. *Physical Review Letters* **2007,** *98* (18), 186806.
23. Tan, Y. W.; Zhang, Y.; Bolotin, K.; Zhao, Y.; Adam, S.; Hwang, E. H.; Das Sarma, S.; Stormer, H. L.; Kim, P., Measurement of Scattering Rate and Minimum Conductivity in Graphene. *Physical Review Letters* **2007,** *99* (24), 246803.
24. Wehling, T. O.; Yuan, S.; Lichtenstein, A. I.; Geim, A. K.; Katsnelson, M. I., Resonant Scattering

by Realistic Impurities in Graphene. *Physical Review Letters* **2010,** *105* (5), 056802.

25. Li, C.; Komatsu, K.; Bertrand, S.; Clavé, G.; Campidelli, S.; Filoramo, A.; Guéron, S.; Bouchiat, H., Signature of gate-tunable magnetism in graphene grafted with Pt-porphyrins. *Physical Review B* **2016,** *93* (4), 045403.

26. Abanin, D. A.; Morozov, S. V.; Ponomarenko, L. A.; Gorbachev, R. V.; Mayorov, A. S.; Katsnelson, M. I.; Watanabe, K.; Taniguchi, T.; Novoselov, K. S.; Levitov, L. S.; Geim, A. K., Giant Nonlocality Near the Dirac Point in Graphene. **2011,** *332* (6027), 328-330.

27. Avsar, A.; Tan, J. Y.; Taychatanapat, T.; Balakrishnan, J.; Koon, G. K. W.; Yeo, Y.; Lahiri, J.; Carvalho, A.; Rodin, A. S.; O'Farrell, E. C. T.; Eda, G.; Castro Neto, A. H.; Özyilmaz, B., Spin–orbit proximity effect in graphene. *Nature Communications* **2014,** *5* (1), 4875.

28. Wojtaszek, M.; Vera-Marun, I. J.; Maassen, T.; van Wees, B. J., Enhancement of spin relaxation time in hydrogenated graphene spin-valve devices. *Physical Review B* **2013,** *87* (8), 081402.

29. Nomura, K.; MacDonald, A. H., Quantum Hall Ferromagnetism in Graphene. *Physical Review Letters* **2006,** *96* (25), 256602.

30. Kharitonov, M., Edge excitations of the canted antiferromagnetic phase of the $\ensuremath{\nu}=0$ quantum Hall state in graphene: A simplified analysis. *Physical Review B* **2012,** *86* (7), 075450.

31. Kharitonov, M., Phase diagram for the ν=0 quantum Hall state in monolayer graphene. *Physical Review B* **2012,** *85* (15), 155439.

32. McCann, E., Asymmetry gap in the electronic band structure of bilayer graphene. *Physical Review B* **2006,** *74* (16), 161403.

33. Hong, X.; Zou, K.; Zhu, J., Quantum scattering time and its implications on scattering sources in graphene. *Physical Review B* **2009,** *80* (24), 241415.

34. Wang, J. I. J.; Yang, Y.; Chen, Y.-A.; Watanabe, K.; Taniguchi, T.; Churchill, H. O. H.; Jarillo-Herrero, P., Electronic Transport of Encapsulated Graphene and WSe2 Devices Fabricated by Pick-up of Prepatterned hBN. *Nano Letters* **2015,** *15* (3), 1898-1903.

35. Dean, C. R.; Young, A. F.; Cadden-Zimansky, P.; Wang, L.; Ren, H.; Watanabe, K.; Taniguchi, T.; Kim, P.; Hone, J.; Shepard, K. L., Multicomponent fractional quantum Hall effect in graphene. *Nature Physics* **2011,** *7* (9), 693-696.

36. Pan, H.; Wang, Q.; Wu, X.; Song, T.; Song, Q.; Wang, J., Thermal annealing effect on the electrical quality of graphene/hexagonal boron nitride heterostructure devices. *Nanotechnology* **2020,** *31* (35), 355001.

# Figures

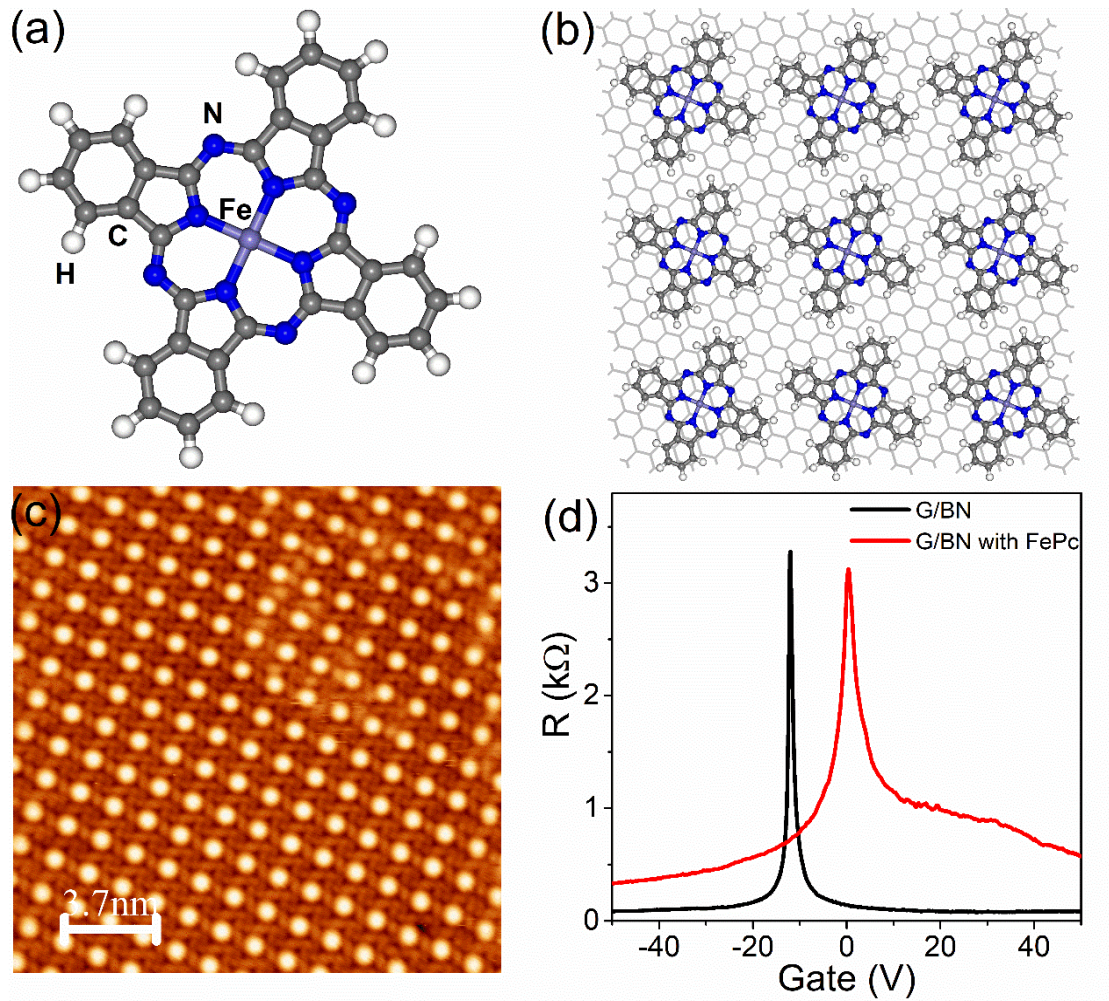

**Figure 1.** FePc molecule structure and characterization. (a) Structure of FePc molecule. (b) DFT calculated structure of FePc molecules on graphene surface. (c) A Scanning tunneling microscopy (STM) image (Tunneling parameters: 200 mV, 0.52 nA) of the FePc molecules adsorbed in a squared closed-packed geometry on the graphite surface. The scale bar is 3.7 nm. (d) Resistance versus applied back gate voltage for monolayer graphene before and after FePc molecules deposition measured at 1.6 K.

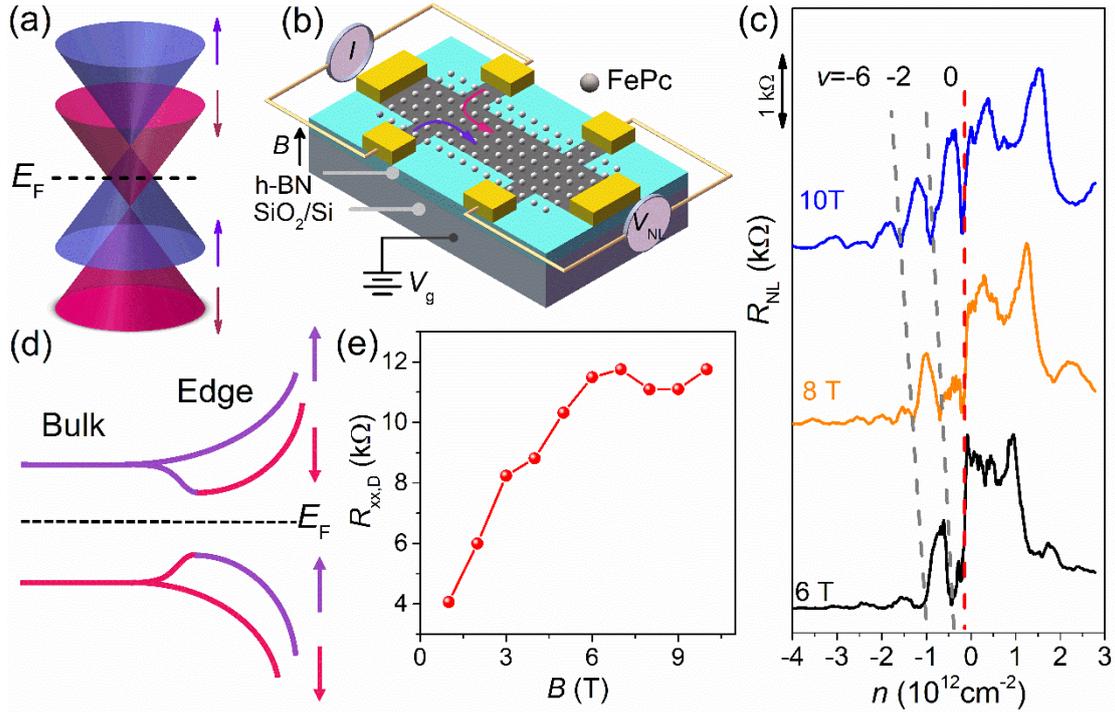

**Figure 2.** Electrical transport of monolayer graphene with FePc molecules at 1.6 K. (a) The schematic of the Zeeman splitting of the Dirac cone in monolayer graphene. (b) The electrodes configuration schematic of non-local measurements. (c) Non-local curves of monolayer G/BN device with FePc molecular film measured at 6 T, 8 T and 10 T. (d) Simplified schematic of the canted antiferromagnetic phase with a gap opening on the helical edge states, where the spectrum contains four branches related to the spin and valley degeneracies. The arrows with different colors (red and purple) denote the lifting of spin degeneracy. (e). The longitudinal resistance $R_{xx,D}$ at the Dirac point versus applied magnetic field.

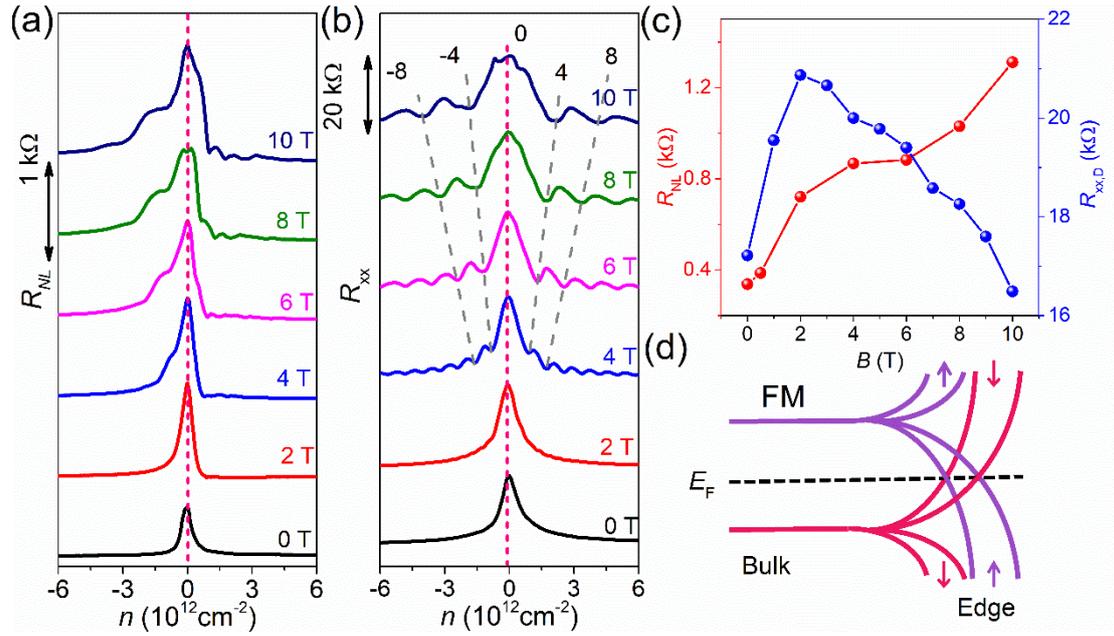

**Figure 3.** Zeeman spin Hall effect in bilayer G/BN with FePc film. (a) The non-local $R_{NL}$ and (b) longitudinal resistance $R_{xx}$ versus carrier density at different fields. All measurements were performed at 1.6 K. (c) The amplitude of $R_{NL}$ and $R_{xx}$ at Dirac point versus magnetic field. (d) Simplified schematics of the ferromagnetism (FM) state in bilayer graphene. The spectrum contains eight branches related to the spin, valley, and orbital degeneracies. The lifting of spin degeneracy is denoted as purple and red colors. The edge is in a gapless state with counter-propagating spin-polarized edge channels.

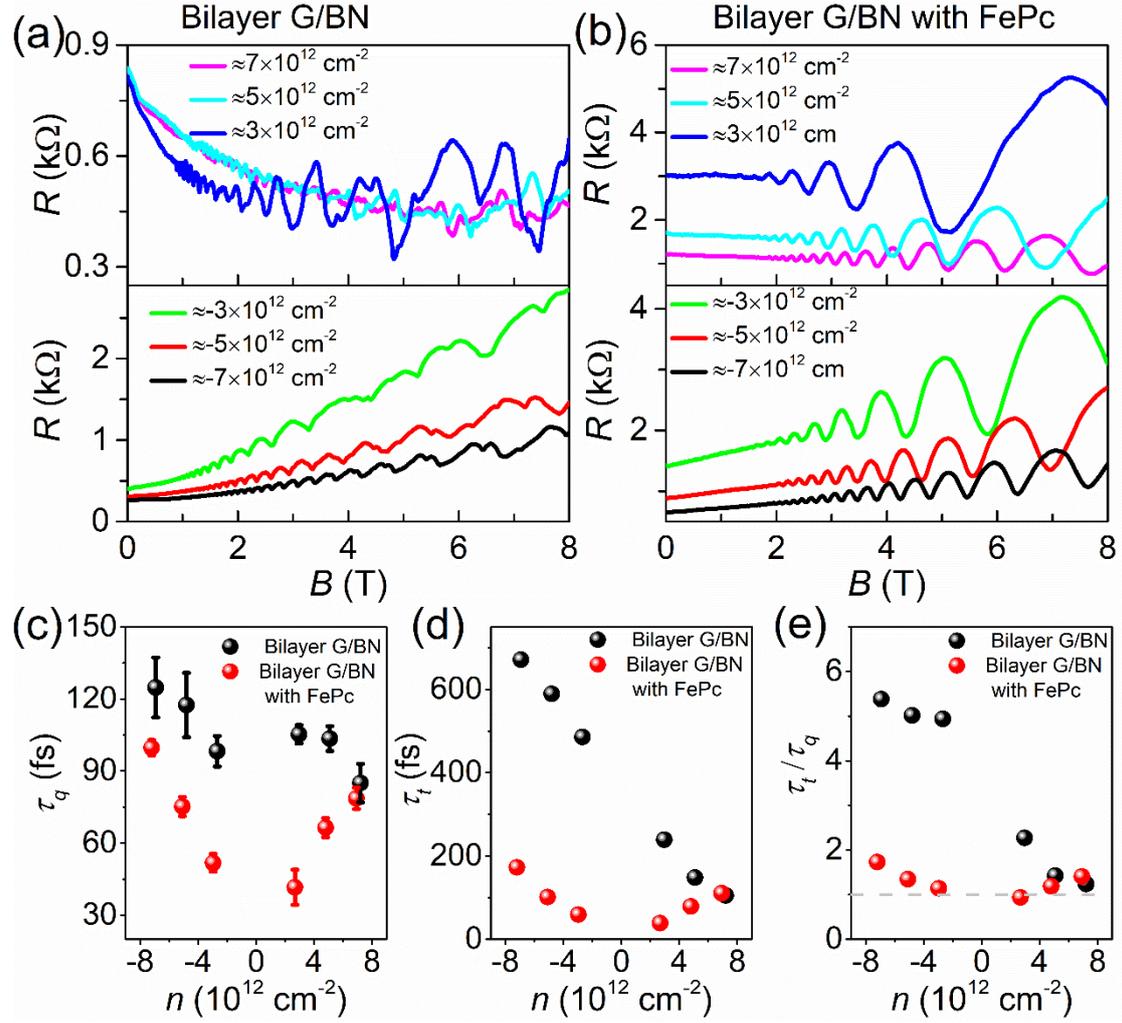

**Figure 4.** SdH oscillation and scattering mechanism analysis of the bilayer G/BN device before and after FePc deposition. (a-b) The magnetoresistance of the bilayer device without (a) and with (b) FePc film measured at various carrier densities, at T = 1.6 K. (c) Quantum scattering time ($\tau_q$) of the bilayer device at various carrier densities without and with FePc film. (d) Transport scattering time ($\tau_t$) of the bilayer device without and with FePc film. (e) The ratio of $\tau_t/\tau_q$ versus carrier density without and with FePc film.